\documentclass[11pt]{article}
\usepackage{amssymb}

\newcommand{\BE}{\begin{equation}}
\newcommand{\EE}{\end{equation}}
\newcommand{\BA}{\begin{eqnarray}}
\newcommand{\EA}{\end{eqnarray}}

\begin{document}
\begin{titlepage}

\vspace*{1mm}
\begin{center}

            {\LARGE{\bf Relativistic analysis of Michelson-Morley 
\\experiments and Miller's cosmic solution 
\\for the Earth's motion }}

\vspace*{14mm}
{\Large  M. Consoli }
\vspace*{4mm}\\
{\large
Istituto Nazionale di Fisica Nucleare, Sezione di Catania \\
c/o Dipartimento di Fisica dell' Universit\`a \\
Via Santa Sofia 64, 95123 Catania, Italy}
\end{center}
\begin{center}
{\bf Abstract}
\end{center}
A simple relativistic
treatment of Michelson-Morley type of experiments
shows the remarkable internal consistency of 1932 Miller's cosmic solution 
$v_{\rm earth} \sim 208$ km/s deduced from the experimental fringe
shifts observed with his apparatus. 
The same analysis of present-day experiments is in 
good agreement with the existing data.
\vskip 35 pt
\end{titlepage}


{\bf 1.}~The aim of this Letter is to illustrate the remarkable
internal consistency of 
1932 Miller's cosmic solution for the Earth's motion. This was deduced from 
the data obtained with
his Michelson-Morley interferometer \cite{miller} on the base of 
the theory exposed by Nassau and Morse \cite{morse}. In this
way, from the variations of the magnitude and azimuth of the ether-drift
effect with the sidereal time, one can determine the apex \cite{apex} 
of the motion of the
solar system. By requiring consistency among the determinations obtained in
four epochs of the year (see Fig.23 of ref.\cite{miller}), 
 Miller obtained values for the Earth's velocity lying in the range 200-215
km/s (see page 233 of ref.\cite{miller}) with the conclusion that 
"...a velocity 
\BE
\label{vcosmic}
v_{\rm earth} \sim 208 ~ {\rm km/s}
\EE
for the cosmic component, gives the closest 
grouping of the four independently determined locations of the cosmic 
apex". 

In this context, it is essential to stress that the fringe shifts observed in 
the classical experiment of Michelson-Morley \cite{mm} (and in the subsequent
one of Morley and Miller \cite{morley}) 
although smaller than the expected magnitude corresponding to the orbital
motion of the Earth, were not negligibly small. While this had already been 
pointed out by Hicks \cite{hicks}, Miller's refined 
analysis of the half-period, 
second-harmonic effect observed in the experimental fringe shifts showed that
they were consistent with an
effective velocity lying in the range 7-10 km/s
(see Fig.4 of ref.\cite{miller}). 
For instance, the Michelson-Morley experiment gave a value
$\sim 8.8 $ km/s for the noon observations 
and a value $\sim 8.0 $ km/s for the evening observations.
By including Miller's Mount Wilson 
results, whose typical effective velocities were also
lying in the range 7-11 km/s (see Fig. 22 of ref.\cite{miller}), one
concludes that all classical ether-drift experiments were consistent with 
an `observable' velocity
\BE
\label{vobs0}
v_{\rm obs} \sim 9 \pm 2~ {\rm km/s. }
\EE
The problem with Miller's analysis was to understand the large discrepancy
between Eq.(\ref{vcosmic}), as needed to describe the {\it variations} of the
ether-drift effect at different sidereal times, and Eq.(\ref{vobs0}), 
 as determined by the magnitude of the fringe shifts themselves.

It has been recently 
pointed out by Cahill and Kitto \cite{cahill} that an effective
reduction of the Earth's velocity from values
$v_{\rm earth}={\cal O}(10^2)$ km/s down to values
$v_{\rm obs}={\cal O}(1)$ km/s can be understood by
taking into
account the effects of the Lorentz contraction and of the
refractive index ${\cal N}_{\rm medium}$ of the
dielectric medium used in the interferometer. 

In this way, the
observations become consistent \cite{cahill} with values of the
Earth's velocity that are comparable to 
$v_{\rm earth} \sim 369$
km/s as extracted by fitting the COBE data for the cosmic
background radiation \cite{cobe}. The point is that the fringe shifts are
proportional to $ {{v^2_{\rm earth} }\over{c^2}}
 (1- {{1}\over{ {\cal N}^2_{\rm medium} }})$ rather than to
                  ${{v^2_{\rm earth} }\over{c^2}}$ itself. For the air, 
where ${\cal N}_{\rm air}\sim 1.00029$, assuming a value
$v_{\rm earth} \sim 369$ km/s, one expects fringe shifts 
governed by an effective velocity 
$v_{\rm obs}\sim 9 $ km/s consistently with Miller's analysis 
Eq.(\ref{vobs0}) of the classical experiments.

This would also explain why the experiments
of Illingworth \cite{illing} ( performed in an apparatus filled with helium 
where ${\cal N}_{\rm helium}\sim 1.000036$) 
and Joos \cite{joos} (performed in the vacuum where
${\cal N}_{\rm vacuum}\sim 1.00000..$)
were showing smaller fringe shifts and, therefore, 
lower effective velocities.

In the following, I shall re-formulate the
argument using Lorentz transformations. As a matter of fact, in 
this case there is a non-trivial difference of a factor $\sqrt{3}$ that makes 
Miller's solution Eq.(\ref{vcosmic}) entirely consistent with 
Eq.(\ref{vobs0}). 
\vskip 10 pt

{\bf 2.}~As a first step, I'll start from
the idea that light propagates in 
 a medium with refractive index ${\cal N}_{\rm medium} > 1$  and
small Fresnel's drag coefficient
\BE
 k_{\rm medium}=
1- {{1}\over{ {\cal N}^2_{\rm medium} }} \ll 1
\EE
(if the medium is the 
vacuum itself, the physical interpretation of ${\cal N}_{\rm vacuum}$
represents a further step, see refs.\cite{pagano,modern}). Let us
also introduce an isotropical speed of light 
($c=2.9979..10^{10}$ cm/s)
\BE
\label{u}
u\equiv  
{{c}\over{\cal{N}_{\rm medium} }}
\EE
The basic question is to determine experimentally, and
to a high degree of accuracy, 
whether light propagates
isotropically with velocity Eq.(\ref{u}) for an observer $S'$ placed on the
Earth. For instance for the air, where the relevant value is
${\cal N}_{\rm air}=1.00029..$, the isotropical value
${{c}\over{\cal{N}_{\rm air} }}$ is usually determined directly by
measuring the two-way speed of light along various directions. In this way, 
isotropy can be established, at best, at the level $10^{-6}-10^{-7}$. 
If we require, however, 
a higher level of accuracy, say $10^{-9}$, the only way 
to test isotropy is to 
perform a Michelson-Morley type of experiment and look for 
fringe shifts upon rotation of the interferometer. 

Now, if one finds experimentally fringe shifts (and thus 
some non-zero anisotropy), 
one can explore the possibility that this effect is due 
to the Earth's motion with respect to a preferred frame 
$\Sigma\neq S'$. In this perspective, light would 
propagate isotropically with velocity as in Eq.(\ref{u})
for $\Sigma$ but {\it not} for $S'$. 

Assuming this scenario, the degree of anisotropy for $S'$ 
can easily be determined by using Lorentz
transformations. By defining
${\bf{v}}$ the velocity of
$S'$ with respect to $\Sigma$ one finds 
($\gamma= 1/\sqrt{ 1- {{ {\bf{v}}^2}\over{c^2}} }$) 
\BE
\label{uprime}
  {\bf{u}}'= {{  {\bf{u}} - \gamma {\bf{v}} + {\bf{v}}
(\gamma -1) {{ {\bf{v}}\cdot {\bf{u}} }\over{v^2}} }\over{ 
\gamma (1- {{ {\bf{v}}\cdot {\bf{u}} }\over{c^2}} ) }}
\EE
where $v=|{\bf{v}}|$. By keeping terms up 
to second order in $v/u$, one obtains
\BE
  {{ |{\bf{u'}}| }\over{u}}= 1- \alpha {{v}\over{u}} -\beta {{v^2}\over{u^2}}
\EE
where ($\theta$ denotes the angle between ${\bf{v}}$ and ${\bf{u}}$)
\BE
   \alpha = (1-  {{1}\over{ {\cal N}^2_{\rm medium} }} ) \cos \theta + 
{\cal O} ( ({\cal N}^2_{\rm medium}-1)^2 )
\EE
\BE
\beta = (1- {{1}\over{ {\cal N}^2_{\rm medium} }} ) P_2(\cos \theta) +
{\cal O} ( ({\cal N}^2_{\rm medium}-1)^2 )
\EE
with $P_2(\cos \theta) = {{1}\over{2}} (3 \cos^2\theta -1)$.

Finally defining $u'(\theta)= |{\bf{u'}}|$, the two-way speed of light is 
\BE
\label{twoway}
{{\bar{u}'(\theta)}\over{u}}= {{1}\over{u}}~ {{ 2  u'(\theta) u'(\pi + \theta) }\over{ 
u'(\theta) + u'(\pi + \theta) }}= 1- {{v^2}\over{c^2}} ( A + B \sin^2\theta) 
\EE
where 
\BE 
\label{ath}
   A= {\cal N}^2_{\rm medium} -1 + {\cal O} ( ({\cal N}^2_{\rm medium}-1)^2 )
\EE
and 
\BE
\label{BTH}
     B= -{{3}\over{2}} 
({\cal N}^2_{\rm medium} -1 )
+ {\cal O} ( ({\cal N}^2_{\rm medium}-1)^2 )
\EE
The above results will be useful in the following. 

Let us now address the theory of the Michelson-Morley
interferometer by considering
two light beams, say 1 and 2, that for simplicity are chosen 
perpendicular in $\Sigma$
where they propagate along the $x$ and $y$ axis with velocities
$u_x(1)=u_y(2)=
u= {{c}\over{ {\cal N}_{\rm medium} }}$.
Let us also assume that the velocity $v$ of  $S'$ is along the $x$ axis.
 In this case, to evaluate the velocities of 1 and 2 for
$S'$, we can apply Lorentz transformations with the result
\BE
 u'_x( 1)= {{ u -v }\over{ 1 - {{uv}\over{c^2}} }}~~~~~~~~~~~~~~u'_y(1)=0
\EE
and
\BE
 u'_x( 2)= -v~~~~~~~~~~~~~~u'_y(2)= u \sqrt{  1- {{v^2}\over{c^2}} }
\EE
Let us now define $L'_P$ and $L'_Q$ to be the lengths of two
optical paths, say P and Q, as
measured in the $S'$ frame. For instance, they can represent the
lengths
of the arms of an interferometer which is at rest in the $S'$ frame.
In the first experimental set-up, the arm
of length $L'_P$ is taken along the direction of motion associated with the beam
1 while the arm of length $L'_Q$ lies along the direction of the beam 2.
Notice that the two arms, in the $S'$ frame, form an angle
that differs from $90^o$ by ${\cal O}(v/c)$ terms.

Therefore, using the above results,
the time for the beam 1 to go forth and back along $L'_P$ is
\BE
          T'_P= L'_P (
{{1- uv/c^2 }\over{u-v}} + {{1+ uv/c^2 }\over{u+ v}} )
\sim
{{2L'_P}\over{u}}
( 1+ k_{\rm medium} {{v^2}\over{u^2}} )
\EE
To evaluate the time $T'_Q$,
for the beam 2 to go forth and back along the arm of length
$L'_Q$, one has first to compute
the modulus of its velocity in the $S'$ frame
\BE
  u'(2) \equiv  \sqrt{
(u'_x(2))^2 + ((u'_y(2))^2 } = u \sqrt{ 1 + k_{\rm medium}{{ v^2}\over{u^2}} }
\EE
and then use the relation $u'(2) T'_Q= 2 L'_Q$ thus obtaining
\BE
  T'_Q = {{2 L'_Q}\over{u'(2)}} \sim
{{2 L'_Q}\over{u}}
( 1- k_{\rm medium} {{v^2}\over{2 u^2}} )
\EE
In this way, the interference pattern, between the light beam coming out
of the optical path P and that coming out of the optical path Q, is
determined by the delay time
\BE
   \Delta T'= T'_P- T'_Q \sim
{{2L'_P}\over{u}}
( 1+ k_{\rm medium} {{v^2}\over{u^2}} )-
{{2 L'_Q}\over{u}}
( 1- k_{\rm medium} {{v^2}\over{2 u^2}} )
\EE
On the other hand, if the beam 2 were to propagate along the optical path
P and
the beam 1 along Q, one would obtain a different
delay time, namely
\BE
   (\Delta T')_{\rm rot}= (T'_P- T'_Q)_{\rm rot}\sim
{{2L'_P}\over{u}} ( 1- k_{\rm medium} {{v^2}\over{2u^2}} )- {{2
L'_Q}\over{u}} ( 1+ k_{\rm medium} {{v^2}\over{u^2}} ) \EE so
that, by rotating the apparatus, there will be a fringe shift
proportional to 
\BE 
\label{deltat} \Delta T'- (\Delta T')_{\rm
rot} \sim {{3(L'_P+ L'_Q)}\over{u}}  k_{\rm medium}
{{v^2}\over{u^2}} 
\EE 
This coincides with the pre-relativistic expression provided one replaces
$v$ with an effective observable velocity
\BE
\label{vobs}
           v_{\rm obs}= v
\sqrt { k_{\rm medium} }
 \sqrt{3}
\EE
Eq.(\ref{deltat}) can also be obtained by using the
equivalent form of the Robertson-Mansouri-Sexl parametrization
\cite{robertson,mansouri} for the two-way speed of light 
defined above in Eq.(\ref{twoway}). In fact, using the relations
\BE
   \Delta T'=
{{2L'_P}\over{\bar{u'}(0)}}- {{2 L'_Q}\over{\bar {u'}(\pi/2)}} \EE
\BE
   {(\Delta T')}_{\rm rot}=
{{2L'_P}\over{\bar{u'}(\pi/2)}}- {{2 L'_Q}\over{\bar {u'}(0)}} \EE
and Eqs.(\ref{ath}) and (\ref{BTH}), 
one obtains 
\BE 
\label{deltat2} \Delta T'- (\Delta T')_{\rm
rot} \sim (-2B) {{(L'_P+ L'_Q)}\over{u}}  
{{v^2}\over{u^2}} 
\EE 
that agrees with Eq.(\ref{deltat}) up to ${\cal O}(k^2_{\rm
medium})$ terms.
 \vskip 10 pt
{\bf 3.}~Now, upon operation of the interferometer, one is faced 
 with several alternatives:

~~~~i) there are no fringe shifts at all. This corresponds to the usual
point of view that light propagates isotropically on the Earth so that
$S'\equiv \Sigma$

~~~ii) there are fringe shifts but their magnitude turns out to
be unrelated to any meaningful definition of the Earth's velocity (
think for instance of some anisotropy
due to the Earth's magnetic field)

~~iii) there are fringe shifts  and their magnitude,
 observed with different dielectric media and within the
experimental errors, points
consistently to a unique value of the Earth's velocity

Case iii) would represent experimental evidence for
the existence of a preferred frame $\Sigma \neq S'$. In practice, to
 ${\cal O}({{v^2_{\rm earth} }\over{c^2}} )$, this can be decided by
re-analyzing \cite{cahill}
the experiments in terms of the effective parameter
 $\epsilon = {{v^2_{\rm earth} }\over{u^2}} k_{\rm medium}$. The
conclusion of Cahill and Kitto \cite{cahill} 
is that the classical experiments are consistent with the value
 $v_{\rm earth}\sim 369$ km/s obtained from the COBE data. 

However, in
our expression Eq.(\ref{vobs}) determining the fringe shifts there is a difference
of a factor $\sqrt{3}$ with respect to their result
$v_{\rm obs}=v \sqrt { k_{\rm medium} }$. 
Therefore, using
Eqs.(\ref{vobs0}) and (\ref{vobs}), for ${\cal N}_{\rm air} \sim 1.00029$, 
 the relevant value of the Earth's velocity is {\it not} 
$v_{\rm earth}\sim 369$ km/s but rather
\BE
\label{vearth}
                  v_{\rm earth}= 216 \pm 47 ~{\rm km/s}
\EE
This is completely consistent with the value  
Eq.(\ref{vcosmic}) obtained by Miller 
from the variations of the 
magnitude and of the azimuth of the ether-drift 
effect with the sidereal time on the base of the
theory of Nassau and Morse \cite{morse}. 
\vskip 10 pt
{\bf 4.}
Still today, the original
Michelson-Morley experiment of 1887 \cite{mm} is considered a {\it proof}
that absolute motion cannot be detected. Actually, the results of
that experiment were smaller than expected but non-zero. 
As pointed out by Cahill and Kitto \cite{cahill}, 
the key-ingredient to understand the reduction from values
$v_{\rm earth}={\cal O}(10^2)$ km/s down to values
$v_{\rm obs}={\cal O}(1)$ km/s, 
consists in taking into account the Lorentz contraction and
the refractive index of the dielectric
medium filling the arms of the interferometer. However, a full treatment
on the base of Lorentz transformations introduces a factor $\sqrt{3}$ with
respect to their analysis. As a consequence the relevant value of the 
Earth's velocity is not $v_{\rm earth}\sim 369$ km/s but rather
 $v_{\rm earth}\sim 216$ km/s. This value, which is completely consistent with
Miller's determination Eq.(\ref{vcosmic}), suggests that 
the magnitude of the fringe shifts is determined by the typical
velocity of the solar system within our galaxy and not, for instance, 
by the velocity of the solar system relatively to the
centroid of the Local Group. 
In the latter case, one would get higher values as
$v_{\rm earth}= 300 \pm 25$ km/sec
ref.\cite{M1}, $v_{\rm earth}= 315 \pm 15$ km/sec ref.\cite{V4},
$v_{\rm earth}= 308 \pm 23$ km/sec ref.\cite{Y2}, $v_{\rm earth}=
336 \pm 17$ km/sec ref.\cite{V5}. 

Notice that such ambiguities, say  
$v_{\rm earth}\sim 200, 300, 369,...$ km/s, on the actual value of the
Earth's velocity determining
the fringe shifts, can only be resolved experimentally in view of the
many theoretical uncertainties in the operative
definition of the preferred frame
where light propagates isotropically. At this stage, I believe, one should 
just concentrate on the internal consistency of the various frameworks. 
In this sense, the simple relativistic analysis presented in this Letter shows
that this is certainly true for Miller's 1932 solution. 

I am aware that my conclusion goes against the
widely spread belief that Miller's results were only due to statistical
fluctuations and/or local temperature conditions (see the Abstract of 
ref.\cite{shankland}). However, it is also true that the same
authors of ref.\cite{shankland} were admitting that "...there can be 
little doubt that statistical fluctuations alone  cannot account for the
periodic fringe shifts observed by Miller" 
(see page 171 of ref.\cite{shankland}).  Even more, although "...there is 
obviously considerable scatter in the data at each azimuth position...the
average values...show a marked second harmonic effect"
(see page 171 of ref.\cite{shankland}). In any case, interpreting the observed
effects on the base of the local temperature conditions is certainly not
the only solution since "...we must admit that a direct and general 
quantitative correlation between amplitude and phase of the observed 
second harmonic on the one hand and the thermal conditions in the observation
hut on the other hand could not be established" 
(see page 175 of ref.\cite{shankland}). This unsatisfactory explanation should
instead be compared with the excellent agreement that was obtained by Miller
once the final parameters for the Earth's velocity were plugged in the 
theoretical predictions (see Figs. 26 and 27 of ref.\cite{miller}). 
Finally, it seems appropriate to remark
that Miller's experiments represented the most refined version of
that `interferometric art' initiated by Michelson and Morley. There is
some
inner contradiction in concluding that Miller was simply wrong in 1932 but 
Michelson and Morley, nevertheless, 
performed in 1887 an experiment that changed the 
history of physics. 
\vskip 10 pt

{\bf 5.}~I conclude with a brief comparison with present-day, 
 `vacuum' Michelson-Morley experiments of the type first performed by 
Brillet and Hall \cite{brillet} and more recently by 
M\"uller et al. \cite{muller}. In this case, by definition 
${\cal N}_{\rm vacuum}=1$ so that $v_{\rm obs}=0$ and no anisotropy can 
be detected. However,  as anticipated above, one can explore 
\cite{pagano,modern} the possibility that, even in this case,
 a very small anisotropy might be due to a refractive index 
${\cal N}_{\rm vacuum}$ that differs from unity by an infinitesimal
amount. In this case, the natural candidate to explain a value
${\cal N}_{\rm vacuum} \neq 1$
is gravity. In fact, by using
the Equivalence Principle, any freely falling frame $S'$ will locally 
measure the same speed of light as in an inertial frame in the absence of
any gravitational effects. However, if $S'$ carries on board an heavy 
object this is no longer true. For an observer placed on the Earth, 
this amounts to insert
the Earth's gravitational potential in the  weak-field isotropic
approximation to the line element of
 General Relativity \cite{weinberg}
\BE
ds^2= (1+ 2\varphi) dt^2 - (1-2\varphi)(dx^2 +dy^2 +dz^2)
\EE
so that one obtains a refractive index for
light propagation 
\BE
\label{nphi}
            {\cal N}_{\rm vacuum}\sim  1- 2\varphi
\EE
This represents the `vacuum' analogue of 
${\cal N}_{\rm air}$, ${\cal N}_{\rm helium}$,...so that from
\BE 
     \varphi =- {{G_N M_{\rm earth}}\over{c^2 R_{\rm earth} }} \sim
-0.7\cdot 10^{-9}
\EE
and using Eq.(\ref{BTH}) one predicts
\BE
\label{theor}
                 B_{\rm vacuum} \sim -4.2 \cdot 10^{-9}
\EE
For $v_{\rm earth} \sim 208$ km/s, 
this implies an observable anisotropy of the two-way speed of light 
in the vacuum Eq.(\ref{twoway}) 
\BE
        {{ \Delta \bar{c}_\theta }\over{c}} \sim 
|B_{\rm vacuum}| {{v^2_{\rm earth} }\over{c^2}} \sim 2\cdot 10^{-15}
\EE
This prediction is in good agreement with 
the experimental value
        ${{ \Delta \bar{c}_\theta }\over{c}}= (2.6 \pm 1.7) \cdot 10^{-15}$ 
determined by M\"ueller et al.\cite{muller}. Notice that the anisotropy
experiment
is sensitive to the product $ B {{v^2_{\rm earth} }\over{c^2}}$ while the
extraction of $B$ from the data
was performed \cite{muller} assuming 
the fixed value $v_{\rm earth}=369$
km/s. Therefore, their determination 
$ B^{\rm exp}= (-2.2 \pm 1.5)\cdot 10^{-9}$,
in addition to the purely experimental error, contains a
{\it theoretical} uncertainty due to
the rigid identification of the cosmic 
background radiation with the preferred
frame where light propagates isotropically. This uncertainty
represents a kind of systematic error whose magnitude can be estimated 
by comparing with alternative definitions 
of the Earth's velocity. For instance, using the 
alternative value 
$v_{\rm earth} \sim 208$ km/s, the same experimental data would produce a
value
$ B^{\rm exp}= (-7.2 \pm 4.9)\cdot 10^{-9}$ that is certainly
consistent with the prediction in Eq.(\ref{theor}).

\vskip 30 pt {\bf Acknowledgements} \vskip 5 pt I thank A. Pagano and
V. Rychkov for useful discussions.


\vskip 60 pt

\begin{thebibliography} {99}
 \bibitem{miller}
D. C. Miller, Rev. Mod. Phys. {\bf 5} (1933) 203.
 \bibitem{morse}
J. J. Nassau and P. M. Morse, ApJ. {\bf 65} (1927) 73.
 \bibitem{apex}
The apex indicates the point on the celestial sphere towards which the Earth is
moving because of its absolute motion.
 \bibitem{mm}
A. A. Michelson and E. W. Morley, Am. J. Sci. {\bf 34} (1887) 333.
 \bibitem{morley}
E. W. Morley and D. C. Miller, Phil. Mag. {\bf 9} (1905) 669.
 \bibitem{hicks}
W. M. Hicks, Phil. Mag. 
{\bf 3} (1902) 9; {\it ibidem} {\bf 3} (1902) 256; 
{\it ibidem} {\bf 3} (1902) 555. 
 \bibitem{cahill}
R. T. Cahill and K. Kitto, arXiV:physics/0205070.
 \bibitem{cobe}
C. H. Lineweaver et al., ApJ. {\bf 470} (1996) 38.
 \bibitem{illing}
K. K. Illingworth, Phys. Rev. {\bf 30} (1927) 692. 
 \bibitem{joos}
G. Joos, Ann. d. Physik {\bf 7} (1930) 385.
 \bibitem{pagano}
M. Consoli, A. Pagano and L. Pappalardo, arXiv:physics/0306094, revised version,
to appear in Physics Letters {\bf A}.
 \bibitem{modern}
M. Consoli, arXiv:gr-qc/0306105.
 \bibitem{robertson}
H. P. Robertson, Rev. Mod. Phys. {\bf 21} (1949) 378.
 \bibitem{mansouri}
 R. M. Mansouri and R. U. Sexl, Gen. Rel. Grav. {\bf 8} (1977) 497.
 \bibitem{M1}
N. U. Mayall, ApJ. {\bf 104} (1946) 290.
 \bibitem{V4}
G. De Vaucouleurs  and W. L. Peters, Nature {\bf 220} (1968) 868.
 \bibitem{Y2}
A. Yahil, A. Sandage and G. A. Tammann, ApJ. {\bf 217} (1977) 903.
 \bibitem{V5}
G. De Vaucouleurs  and W. L. Peters, ApJ. {\bf 248} (1981) 395.
 \bibitem{shankland}
R. S. Shankland, S. W. McCuskey, F. C. Leone, and G. Kuerti, Rev. Mod. Phys. 
{\bf 27} (1955) 167. 
 \bibitem{brillet}
A. Brillet and J. L. Hall, Phys. Rev. Lett. {\bf 42} (1979) 549.
 \bibitem{muller}
H. M\"uller, S. Herrmann, C. Braxmaier, S. Schiller and A. Peters,
arXiv:physics/0305117.
 \bibitem{weinberg}
S. Weinberg, {\it Gravitation and Cosmology}, John Wiley and Sons, Inc., 1972,
pag. 181.


\end{thebibliography}
\end{document}